\documentclass[11pt]{article}
\usepackage[title]{appendix}
\usepackage[margin={1.0in}]{geometry}
\usepackage{rotating, lscape}
\usepackage{float}
\usepackage{graphicx, color}

\usepackage{threeparttable, booktabs, longtable, threeparttablex, tabularx, multirow, pdflscape}

 \usepackage{natbib}
\usepackage[sc]{mathpazo}
\usepackage{courier}
\usepackage[utf8]{inputenc}
\usepackage[T1]{fontenc}
\usepackage{setspace} 

\usepackage{amssymb,amsmath,mathrsfs,amsthm, bbm}
\usepackage{accents}

\usepackage{microtype}
\usepackage[colorlinks=true, citecolor=black, urlcolor=blue]{hyperref}
\usepackage{afterpage}
\usepackage{caption}
\usepackage{amsmath}
\usepackage{subfigure}
\usepackage{pdflscape} 
\usepackage{makecell}
\usepackage{multirow}

\RequirePackage[font=small,format=plain,labelfont=bf,textfont=it]{caption}
\usepackage{chngcntr}
\usepackage{palatino}
\usepackage{threeparttable}
\usepackage{amsthm}
\usepackage{ragged2e}
\usepackage{subcaption}
\newtheorem{theorem}{Theorem}

\newtheorem{assumption}{Assumption}
\usepackage{tikz}
\usetikzlibrary{positioning}

\usepackage{textcomp}
\usepackage{tkz-graph}
\usetikzlibrary{shapes.geometric}

\usepackage{hyperref}
\usepackage{xcolor}
\hypersetup{
  colorlinks   = true, 
  urlcolor     = blue, 
  linkcolor    = blue, 
  citecolor   = blue 
}

\usepackage{todonotes}
\newcounter{todocounter}

\newcolumntype{Y}{>{\centering\arraybackslash}X}

\newcolumntype{L}{>{\raggedleft\arraybackslash}X}

\newcolumntype{R}{>{\raggedright\arraybackslash}X}

{\begin{minipage}{\linewidth}%
\smallskip\footnotesize\emph{Notes:}}%
{\end{minipage}}

\usetikzlibrary{shapes,decorations,arrows,calc,arrows.meta,fit,positioning}
\tikzset{
    -Latex,auto,node distance =1 cm and 1 cm,semithick,
    state/.style ={ellipse, draw, minimum width = 0.7 cm},
    point/.style = {circle, draw, inner sep=0.04cm,fill,node contents={}},
    bidirected/.style={Latex-Latex,dashed},
    el/.style = {inner sep=2pt, align=left, sloped}
}

\usepackage{etoolbox}
\makeatletter
\def\@fnsymbol#1{\ifcase#1\or *\or **\or ***\else\@ctrerr\fi}
\makeatother

\title{Learning and Testing Exposure Mappings of Interference using Graph Convolutional Autoencoder}
\author{{Martin Huber}\footnote{University of Fribourg; martin.huber@unifr.ch} \space , 
 {Jannis Kueck}\footnote{Heinrich Heine University Düsseldorf; kueck@dice.hhu.de} \space
 , {Mara Mattes}\footnote{Heinrich Heine University Düsseldorf; mattes@dice.hhu.de}
}
\date{\today}
\begin{document}

\renewcommand{\thefootnote}{\arabic{footnote}}

\maketitle 
\thispagestyle{empty}

\begin{abstract}
\noindent 

  Interference or spillover effects arise when an individual's outcome (e.g., health) is influenced not only by their own treatment (e.g., vaccination) but also by the treatment of others, creating challenges for evaluating treatment effects. Exposure mappings provide a framework to study such interference by explicitly modeling how the treatment statuses of contacts within an individual's network affect their outcome. Most existing research relies on a priori exposure mappings of limited complexity, which may fail to capture the full range of interference effects. In contrast, this study applies a graph convolutional autoencoder to learn exposure mappings in a data-driven way, which exploit dependencies and relations within a network to more accurately capture interference effects. As our main contribution, we introduce a machine learning-based test for the validity of exposure mappings and thus test the identification of the direct effect. In this testing approach, the learned exposure mapping is used as an instrument to test the validity of a simple, user-defined exposure mapping. The test leverages the fact that, if the user-defined exposure mapping is valid (so that all interference operates through it), then the learned exposure mapping is statistically independent of any individual's outcome, conditional on the user-defined exposure mapping. 
  We assess the finite-sample performance of this proposed validity test through a simulation study.

\end{abstract}

\noindent \emph{Keywords:} causal inference, network interference, graph convolutional autoencoder, conditional independence, instrumental variables%

	\vspace{0.3cm} \noindent 
\newpage
\setcounter{page}{1}

\section{Introduction}

Interference or spillover effects, where an individual’s outcome (e.g., health) is influenced not only by their own treatment (e.g., vaccination) but also by the treatment of others, pose substantial challenges for causal inference, particularly when arbitrary forms of interference are allowed; see \citet{Manski2013}. Exposure mappings, as discussed in \citet{AronowSamii2017} and \citet{eckles2017design}, provide a structured framework to study such interference by explicitly modeling the mechanisms through which the treatment statuses of social contacts within an individual's network influence their outcome. For example, an exposure mapping might specify that an interference effect occurs if at least one contact is treated, but does not depend on the number of treated contacts. By defining such mappings, researchers can separate interference effects from the direct effect of a treatment, provided that they appropriately account for differences in the probabilities of specific exposure mappings across individuals - for instance, due to variation in network structure.
However, most existing research relies on exposure mappings of limited complexity that are a priori defined by the researcher in an ad hoc manner, which risks failing to capture the full range of interference effects. In this paper, we apply graph convolutional autoencoder (GCAs), which are a specific type of graph neural networks (GNNs), to learn exposure mappings in a data-driven way based on network embeddings. By exploiting network dependencies and relations, GCAs allow us to capture complex patterns of interference that may be missed by simple, a priori defined mappings. As our main contribution, we introduce a machine learning-based testing framework for the validity of exposure mappings. Our approach uses a learned, complex exposure mapping as an instrument to assess whether a given simpler, researcher-defined exposure mapping sufficiently captures all interference. Specifically, if the simpler mapping accurately captures all interference effects such that any interference operates through them, then the learned mapping should be statistically independent of the individual's outcome, conditional on the less complex mapping. By examining violations of this condition, we can assess whether less complex, researcher-defined mappings fail to capture some of the underlying interference. 
Although developed in the context of interference, our testing framework is related to conditional independence testing approaches in \citet{deLunaJohansson2012} and \citet{huberkueck2022}, who investigate tests for joint satisfaction of conditional treatment exogeneity and instrumental variable (IV) assumptions. Analogously, we test whether the user-defined exposure mapping is both exogenous and sufficient to capture all interference effects. If the exposure mapping is correctly specified, then the treatment assignments of others affect an individual's outcome only through this mapping, implying an IV type exclusion restriction. In this case, the learned embedding should be conditionally independent of the outcome given the user-defined exposure mapping, which forms the basis of our test.
The testing approach is also related to the causal framework of \citet{yao2025third}, who discuss how causal representation learning (such as the learning of exposure mappings from networks) can be interpreted within a measurement model perspective, where the learned representations are viewed as proxy measurements of latent causal variables (e.g., the entire interference network).

We base the estimation of the direct effect of the treatment on the outcome, as well as the proposed testing procedure, on the double machine learning (DML) framework of \citet{Chetal2018}, which builds on doubly robust score functions; see \citet{Robins+94} and \citet{RoRo95}. The DML framework satisfies the \citet{Neyman1959}-orthogonality condition, which implies that the resulting estimators and tests are relatively insensitive to modest approximation errors in the estimation of the exposure mapping and in the estimation of the treatment or outcome models. Consequently, the estimators and tests satisfy asymptotic normality under specific regularity conditions, in particular if the machine learning methods used to estimate these models, such as GNNs, converge at a rate of $o(n^{-1/4})$ to the respective true model.
Our study contributes to the growing literature that seeks to construct exposure mappings by exploiting information contained in the data. For instance, \citet{bargaglistoffi2023heterogeneoustreatmentspillovereffects} propose a tree-based method to assess heterogeneity in direct treatment and interference effects with respect to individual, neighborhood, and network characteristics. In their framework, it is assumed that interference occurs only within, but not across, predefined clusters (e.g., geographic regions) - a setting known as partial interference; see, e.g., \citet{Sobel2006}, \citet{HongRaudenbush2006}, and \citet{HudgensHalloran2008}. However, interference may still vary depending on the network structure within clusters. The proposed method therefore aims at detecting network structures and classifying clusters that exhibit similar interference effects based on tree-based algorithms. In contrast, our method is not confined to learning exposure mappings within clusters. Instead, we learn exposure mappings as embeddings in a GCA, which constitutes a highly flexible approach capable of capturing complex network dependencies. Furthermore, we complement this estimation strategy with a novel statistical testing procedure to validate exposure mappings.
\citet{pmlr-v130-ma21c} also propose a GNN-based approach to learn interference from the data, focusing on the problem of learning optimal treatment policies across subgroups (see, e.g., \cite*{Manski2004}, \cite*{HiranoPorter2008}, \cite*{KitagawaTetenov2018}, \cite*{AtheyWager2018}) in the presence of interference.
In contrast, the focus of our study is not on optimal policy learning but rather testing whether the exposure mapping is correctly specified which allows the identification of the direct treatment effect.
\citet{NEURIPS2019_af1c25e8} consider the estimation of direct treatment effects while controlling for network-related confounders by conditioning on embeddings learned from networks using DML. To this end, they adapt the DML assumptions in \citet{Chetal2018} to account for learned embeddings and present high-level conditions under which estimation of the direct treatment effect is $\sqrt{n}$-consistent. Also \citet{10.1145/3534678.3539299} discuss direct effect estimation when accounting for confounding induced by network interference. 
More closely related to our setting, \citet{leung2022graph} propose using GNNs to control for network-induced confounding, with the goal of estimating both direct and interference effects and conducting statistical inference. To this end, they rely on the assumption of approximate neighborhood interference (ANI) introduced by \cite*{leung2022causal}, which is conceptually related to approximate sparsity as considered in lasso regression by \citet{Bellonietal2014}. ANI posits that interference decays sufficiently fast with the distance between individuals in the network, thereby addressing the problem of potentially high-dimensional confounding induced by network structure. \citet{leung2022graph} show that, under ANI and additional regularity conditions, the estimation of propensity score and outcome models can achieve convergence at a rate of $o(n^{-1/4})$, implying that direct and interference effect estimators based on the given exposure mappings can be $\sqrt{n}$-consistent and asymptotically normal when implemented within the DML framework. Importantly, the exposure mapping is not learned but prespecified by the researcher and the GNNs are only used to estimate the nuisance functions. These results demonstrate that some restriction on the complexity of interference is necessary to obtain well-behaved estimators. Specifically, the depth of the relevant interference network must not be too large, which allows GNNs with a limited number of layers to approximate the interference structure. \citet{baharan2025graph} also propose a graph-based doubly robust estimator that uses graph neural networks to flexibly learn network confounding and to estimate both direct and interference effects using a prespecified exposure mapping.
An alternative strategy for reducing complexity is proposed by \citet{belloni2022neighborhood}, who permit the depth of the relevant interference network to vary across individuals but, in turn, impose additive separability of the direct and interference effects. Another relevant study in this context is \citet{wang2024graph}, who - albeit not focusing specifically on estimation of interference effects via exposure mappings - establish convergence rates for GNN estimators and derive high-level conditions under which the complexity of network interference (and the approximation error when estimating it) are sufficiently small to attain $o(n^{-1/4})$ rates.
The issue of estimating exposure mappings is also related to the framework of DML estimation with generated (rather than directly observed) regressors, as considered, for instance, in models with estimated control functions; see e.g. \citet{pan2024locally} and \citet{escanciano2023automatic}.
Our study also applies the DML methodology in the context of interference and GNNs, but extends this line of research by proposing a statistical test to assess the validity of the exposure mappings.

The remainder of this study is organized as follows. Section \ref{assumptions} introduces the causal framework, including the concept of exposure mappings, the causal parameters of interest - namely the direct, interference and total effects - and the identifying assumptions. Section \ref{exposuremappings} describes how exposure mappings can be learned from the data using GCAs. Section \ref{testability} presents the testability conditions for exposure mappings and outlines the implementation of corresponding test using DML. Section \ref{estimation} details the DML-based estimation of the causal parameters under correct specification of the exposure mapping. Section \ref{sim} presents a simulation study investigating the finite-sample performance of the estimators of causal effects, as well as of the testing procedure. Section \ref{conclusion} concludes.

\section{Causal effects and identification}\label{assumptions}

This section introduces the concept of exposure mappings, as well as the definition and identification of direct, interference and total effects under specific assumptions. Let $D_i$ denote the treatment of individual $i$ in a population of interest, and let $\mathcal{D}_{-i}$ represent the vector of treatments assigned to all other individuals (excluding individual $i$) in that population. Furthermore, let $Y_i$ denote the observed outcome. Throughout, we use uppercase letters to denote random variables and lowercase letters for specific realizations. Using the potential outcomes framework, as proposed by \cite*{Neyman23} and advocated by \cite*{Rubin74}, the potential outcome of individual $i$ under specific treatment assignments $D_i = d$ and $\mathcal{D}_{-i} = \mathbf{d}$ is written as $Y_i(d, \mathbf{d})$. This contrasts with the standard assumption in most treatment evaluations, which impose the Stable Unit Treatment Value Assumption (SUTVA) \citep{Rubin80, Cox58}. SUTVA rules out interference effects, implying that the potential outcome depends only on individual $i$'s own treatment: $Y_i(d)$. For the subsequent discussion, we assume a binary treatment, such that $d \in \{0,1\}$.

When SUTVA is violated, one pathway to identifying direct, interference, and total effects of an individual's own treatment relies on exposure mappings, see \citet{AronowSamii2017}. Such mappings impose structure on how the treatment assignments of other individuals, $\mathcal{D}_{-i}$, influence the outcome of individual $i$. It is assumed that interference effects operate through an individual's social network, denoted by $\mathcal{A}$, which is observed - for example, as an adjacency matrix indicating which individuals interact with one another.\footnote{The network 
\(\mathcal{A}\) is typically assumed to be fixed. However, \cite*{li2022random} consider the network structure in a population as a random draw and propose an asymptotic framework for constructing confidence intervals for direct and interference effects under this assumption.} Exposure mappings can be viewed as sufficient statistics for capturing any interference effects within a network. More formally, the exposure for individual $i$, denoted by $Z_i$, defines strengths of interference as a function (or mapping) $\mathcal{F}$ of $i$’s network $\mathcal{A}$, the treatment assignments of other individuals, $\mathcal{D}_{-i}$, and the covariates of other individuals $\mathcal{X}_{-i}$:
\begin{eqnarray}\label{exposuremap}  
Z_i=\mathcal{F}(\mathcal{A},\mathcal{D}_{-i},\mathcal{X}_{-i}).
\end{eqnarray} 
The complexity of interference captured by the function $\mathcal{F}$ determines the number of possible values $Z_i$ can take. A common choice in the literature defines $\mathcal{F}$ as the number of individuals who are both treated according to $\mathcal{D}_{-i}$ and neighbors of individual $i$ in the network $\mathcal{A}$, in which case $Z_i$ takes values $z \in \{0,1,2,\ldots\}$. A simpler alternative specifies $\mathcal{F}$ as a binary indicator, where $Z_i = 1$ if at least one treated individual in $\mathcal{D}_{-i}$ is a neighbor of individual $i$ in network $\mathcal{A}$, and $Z_i = 0$ otherwise. This results in a binary exposure: $z \in \{0,1\}$. It is worth noting that we also allow the exposure mapping to depend on the covariates of other individuals.

\begin{figure}[ht!]
\centering
\caption{%
Causal diagram satisfying Assumption \ref{ass1}.
}\label{fig:dag}
\begin{tikzpicture}[transform shape, node distance=1.5cm, thick, roundnode/.style={circle, draw, inner sep=1pt, minimum size=7mm}, squarenode/.style={rectangle, draw, inner sep=1pt, minimum size=7mm}]
\node[roundnode] (D) {$\mathcal{D}_{-i}$};
\node[roundnode, right=of D] (Z) {${Z}_i$};
\node[roundnode, above=of Z] (A) {$\mathcal{A}$};
\node[roundnode, right=of Z] (Y) {$Y_i$};
\node[roundnode, below=of Z] (Di) {$D_i$};
\node[roundnode, right=of Di] (Xi) {$X_i$};
\node[roundnode, left=of A] (X) {$\mathcal{X}_{-i}$};

\draw[-latex] (D) -- (Z);
\draw[-latex] (A) -- (Z);
\draw[-latex] (Z) -- (Y);
\draw[-latex] (Di) -- (Y);
\draw[-latex] (X) -- (D);
\draw[-latex] (Xi) -- (Di);
\draw[-latex] (Xi) -- (Y);
\draw[-latex] (X) -- (Z);
\end{tikzpicture}
\end{figure}
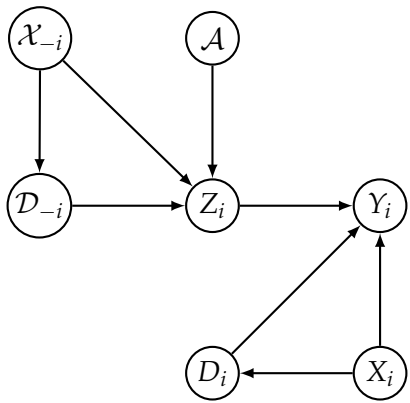

Under a correctly specified exposure mapping in Equation \eqref{exposuremap}, the potential outcome $Y_i(d, \textbf{d})$ simplifies to $Y_i(d, z)$. This permits defining the average direct effect, interference effect, and total effect, denoted by $\gamma(z)$, $\delta(d, z, z')$, and $\Delta(z, z')$, respectively, which are functions of an individual's own treatment and the exposure:
\begin{align} \label{interferenceeff2} \gamma(z)&=E[Y_i(1, z) - Y_i(0, z)],\\ 
\delta(d, z, z')&=E[Y_i(d, z) - Y_i(d, z')],\notag\ \\
\Delta(z, z')&=E[Y_i(1, z) - Y_i(0, z')],\notag \end{align} 
where $z$ and $z'$ are two distinct exposures (e.g.,\ 1 and 0 in the binary case). Next, we provide the formal assumptions on the causal structure that ensure identification of the causal parameters defined in Equation \eqref{interferenceeff2}. We assume that we observe data $W_i=(Y_i,D_i, X_i)$ for a fixed network $\mathcal{A}$, $i=1,\dots,n$, where $(X_i,D_i)$ is $i.i.d.$, but $Y_i$ may depend on $\mathcal{D}_{-i}$ and $\mathcal{X}_{-i}$. Our first assumption concerns the exposure mapping and the treatment assignment:

\begin{assumption}[Identification - Independence of $D_i$ and $Z_i$]\label{ass1}
\begin{align*}
Z_i=\mathcal{F}(\mathcal{A},\mathcal{D}_{-i},\mathcal{X}_{-i}),\ \text{and}\ D_i=D_i(X_i).
\end{align*}
\end{assumption}
First, Assumption \ref{ass1} states that the true exposure $Z_i=\mathcal{F}(\mathcal{A},\mathcal{D}_{-i},\mathcal{X}_{-i})$ is a function only of $\mathcal{A}$, $\mathcal{D}_{-i}$, and $\mathcal{X}_{-i}$. Second, the treatment assignment of individual $i$ may depend only on its own characteristics $X_i$, that is, $D_i=D_i(X_i)$.
This implies that individual $i$'s treatment assignment $D_i$ is conditionally independent of its own exposure $Z_i$ given the covariates $\mathcal{X}=(X_i,\mathcal{X}_{-i})$ and the network structure $\mathcal{A}$. 
As discussed in \cite*{AronowSamii2017}, the propensity score under inference is defined as the joint conditional probability of the treatment and the exposure, given the network structure $\mathcal{A}$ and the observed covariates $\mathcal{X}=(X_i,\mathcal{X}_{-i})$. Under Assumption \ref{ass1}, the propensity score is given by
\begin{align}\label{pscore}
p_i(d, z) = \Pr(D_i = d, Z_i = z | \mathcal{X}, \mathcal{A}) = \Pr(D_i = d | X_i) \cdot \Pr(Z_i = z | \mathcal{X}_{-i}, \mathcal{A}).
\end{align}
Using a directed acyclic graph (DAG) (see, e.g. \cite{Pearl00}), Figure \ref{fig:dag} represents a causal structure where Assumption \ref{ass1} is satisfied. In the graph, nodes represent variables, and arrows indicate causal associations between those variables. The treatment assignments of other individuals \(\mathcal{D}_{-i}\), the network structure \(\mathcal{A}\) and their covariates \(\mathcal{X}_{-i}\) jointly determine the exposure \(Z_i\), which influences the outcome \(Y_i\) in the presence of interference. The confounder $X_i$ influences individual treatment assignment \(D_i\) and the outcome \(Y_i\), whereas $\mathcal{X}_{-i}$ influences the treatment assignments of other individuals \(\mathcal{D}_{-i}\). 
In this structure, the direct effect refers to the effect of an individual's own treatment $D_i$ on their outcome $Y_i$, i.e., $D_i \rightarrow Y_i$. The interference effect corresponds to the impact of other individuals' treatment assignments $\mathcal{D}_{-i}$ on individual $i$'s outcome $Y_i$ through the exposure $Z_i$, i.e., $\mathcal{D}_{-i} \rightarrow Z_i \rightarrow Y_i$. We assume that both the network structure \(\mathcal{A}\) and the covariates of other individuals $\mathcal{X}_{-i}$ do not directly affect the individual $i$'s outcome $Y_i$, but only through $Z_i$. 

Consistent with the causal structure in Figure \ref{fig:dag}, identification of the direct treatment effect of $D_i$ requires that all backdoor paths from $D_i$ to $Y_i$ are blocked by the individual $i$'s observed covariates $X_i$ and the exposure $Z_i$. This motivates the following conditional exogeneity assumption:
\begin{assumption}[Identification - Conditional exogeneity of $D_i$ and $Z_i$]\label{ass2}
\begin{align*}
Y_i(d,z) \perp\!\!\!\perp (D_i,Z_i) | \mathcal{X},\mathcal{A} \quad \forall d \in \{0,1\},\ z\in\mathcal{Z}.
\end{align*}
where $\mathcal{Z}$ denotes the  support of $Z_i$.
\end{assumption}

Assumption \ref{ass2} imposes that there are no unobserved variables that jointly affect $Y_i$ and the treatment assignment $D_i$, or $Y_i$ and the true exposure $Z_i$ conditional on $\mathcal{X}$ and $\mathcal{A}$. Notably, under our assumed causal structure, there is no direct effect of $\mathcal{A}$ or $\mathcal{X}_{-i}$ on the outcome $Y_i$. As a result, conditioning on $X_i$ alone would be sufficient in Assumption \ref{ass2}, since all paths from $\mathcal{X}_{-i}$ and $\mathcal{A}$ to $Y_i$ are blocked by $Z_i$. 

Furthermore, we require that the propensity score for any combination of treatment and exposure is strictly positive. This implies that the network structure does not deterministically determine the exposure. Thus, there exists variation in exposures, conditional on the network and the covariates, that can be leveraged to assess their effects. This leads to the following common support assumption:
\begin{assumption}[Identification - Common support]\label{ass3}
\begin{align*}
p_i(d, z) > 0 \quad \forall d \in \{0,1\}, z \in \mathcal{Z}.
\end{align*}
\end{assumption}

Under Assumptions \ref{ass1} - \ref{ass3}, the causal effects defined in Equations \eqref{interferenceeff2} are identified through the propensity score. As discussed in \cite*{aronow2020spillover}, inverse probability weighting (IPW) \citep{Horvitz52} can be applied, reweighting observations by the inverse of the propensity score to recover the mean potential outcomes and effects:
\begin{align}\label{ipwinterference}
E[Y_i(d,z)]&=E \left[ \frac{Y_i\cdot I \{D_i=d, Z_i=z\}}{p_i(d,z)} \right],\\
\gamma(z)&= E \left[ \frac{Y_i\cdot D_i \cdot I \{Z_i=z\}}{p_i(1,z)} - \frac{Y_i\cdot (1-D_i) \cdot I \{Z_i=z\}}{p_i(0,z)}\right],\notag\\
\delta(d,z,z')&= E \left[ \frac{Y_i\cdot I \{D_i=d, Z_i=z\}}{p_i(d,z)} - \frac{Y_i\cdot I \{D_i=d, Z_i=z'\}}{p_i(d,z')} \right],\notag\\
\Delta(z,z')&= E \left[ \frac{Y_i\cdot D_i \cdot I \{Z_i=z\}}{p_i(1,z)} - \frac{Y_i(1-D_i)  \cdot I \{Z_i=z'\}}{p_i(0,z')} 
\right],\notag
\end{align}
where $I \{\cdot \}$ denotes the indicator function, which is one if its argument is satisfied and zero otherwise. 

\section{Learning exposure mappings}\label{exposuremappings}

While most studies evaluating interference effects rely on researcher-specified mappings, in real-world applications the function $\mathcal{F}$, and thus the true definition of the exposure mapping, is unknown. For this reason, we aim to approximate the true exposure, $Z_i$, using a graph convolutional autoencoder (GCA) related to the graph autoencoder (GAE) of \citet{kipf2016variationalgraphautoencoders}\footnote{Although inspired by the GAE of \citet{kipf2016variationalgraphautoencoders}, our GCA differs in that it is trained to predict $Y_i$ in a supervised setting, rather than constructing the adjacency matrix.}. 
In our supervised learning approach, $\tilde{Z}_i$ is learned as the embedding from an intermediate hidden layer of a GCA trained to predict the outcome $Y_i$ from the network $\mathcal{A}$, the treatment assignments of the individuals $\mathcal{D}=(D_i,\mathcal{D}_{-i})$, and the observed covariates $\mathcal{X}=(X_i,\mathcal{X}_{-i})$. A graph autoencoder consists of an encoder that maps the input graph and node features into a low-dimensional latent representation (embedding) capturing its most relevant information, and a decoder that predicts target values from this latent representation. In our setting, the GCA combines an encoder that consists of graph convolutional network (GCN) layers \citep{kipf2017semisupervisedclassificationgraphconvolutional} and a regression-based decoder. This architecture allows the model to automatically learn dependencies and relationships between nodes based on the network structure. The resulting learned exposures, denoted by $\tilde{Z}_i$, correspond to the embeddings produced by the encoder and are designed to capture how treatment assignments across the network affect individual outcomes. 

We consider an undirected graph $G = (\mathcal{V},\mathcal{E})$, where the nodes are individuals $v_i \in \mathcal{V}$ with $|\mathcal{V}| = n$ and the edges are given by pairs $(v_i,v_j) \in \mathcal{E}$. The adjacency matrix $\mathcal{A} \in \mathbb{R}^{n \times n}$ represents the connections between the individuals and is defined as
\begin{equation} 
\mathbf{\mathcal{A}}  = 
\begin{bmatrix}
0 & a_{12} & \dots & a_{1n}  \\
a_{21} & 0 & \dots & a_{2n}  \\
\vdots & \vdots & \ddots & \vdots \\
a_{n1} & a_{n2} & \dots & 0 \\
\end{bmatrix}.
\end{equation}

In the adjacency matrix, $a_{ij} = 1$ indicates that node $i$ (i.e., individual $i$) is connected to node $j$, while $a_{ij} = 0$ indicates that there is no edge, i.e., no social connection, between nodes $i$ and $j$. Each node has its own features and the input feature vector consists of the treatment assignments and the covariates of all individuals. Denote this matrix by $M$ with dimensions $n \times 2$, where $d_i$ is the individual assignment and $x_i$ is the covariate of individual $i$: 
\begin{equation}
\label{eq:input}
M_{n \times 2} = \begin{bmatrix} 
d_{1} & x_{1}  \\
d_{2} & x_{2}   \\
\vdots &  \vdots  \\
d_{n} & x_{n}  \\
\end{bmatrix}.
\end{equation}

Each layer in the encoder is indexed by $k \in \{0,\dots , K-1\}$, where $k=0$ corresponds to the raw input, $k=1$ to the first layer and $K-1$ to the final layer of the encoder. The representation of the raw input (layer $k=0$) is $H^{(0)} = M$. The graph convolutional layers $k \in \{1, \dots , K-1\}$ in the encoder follow the layer-wise propagation rule
\begin{equation}
\label{eq:encoder}
H^{(k+1)} = \sigma \big(T^{-\frac{1}{2}}\mathcal{A}T^{-\frac{1}{2}}H^{(k)}\tilde{W}^{(k)}   \big), \\
\end{equation}

where $T$ is the degree matrix. The matrix $T$ contains zeros everywhere except on the diagonal, where $t_{ii} = \sum_j a_{ij}, \forall i \in \{1, \dots, n\}.$ $\tilde{W}^{(k)}$ is a trainable and layer-specific weight matrix, and $\sigma(\cdot)$ is an activation function. $H^{(k+1)} \in \mathbb{R}^{n \times x_k}$ contains the node representations with each row corresponding to one node, where $x_k$ is the feature dimension of that layer. The final encoder layer $H^{(K)}$, with $k = K-1$, has feature dimension $x_{k}=1$, as its output serves as the learned exposure $\tilde{Z}_i$, which we model as a one-dimensional embedding, that is, $ H^{(k+1)} \in \mathbb{R}^{n \times 1}$ for $k=K-1$. This layer-wise update in the graph convolutional layers corresponds to a standard message-passing mechanism, in which each node aggregates transformed information from its neighbors according to the graph structure in $\mathcal{A}$.
To obtain an outcome prediction $\hat{\mathcal{Y}} \in \mathbb{R}^{n \times 1}$, the embeddings from the encoder are passed into a regression-based decoder implemented as a linear layer, which maps the learned exposure representation into a predicted outcome:
\begin{equation}
\label{eq:decoder}
\hat{\mathcal{Y}} = \tilde{W}_{dec}H^{(K)} + b_{dec}
\end{equation}
where $\tilde{W}_{dec} \in \mathbb{R}^{1 \times 1}$ is the trainable weight in the decoder layer, $H^{(K)}$ is the vector of learned exposures $\tilde{Z}_i$, $i\in\{1\dots,n\}$, and $b_{dec} \in \mathbb{R}$ is the bias term. The architecture of the GCA is illustrated in Appendix \ref{GCA} Figure \ref{fig:architecture}.

The GCA is trained by minimizing the mean squared error loss function $\mathcal{L}(Y_i,\hat{Y}_i) = \frac{1}{n} \sum_i (Y_i-\hat{Y}_i)^2$, which is appropriate, because the outcome variable $Y_i$ is continuous.

It is important to note that the adjacency matrix $\mathcal{A}$ contains no self-loops, i.e., $a_{ii}=0, \forall i \in \{1, \dots, n\}$. As a result, each node aggregates information exclusively from its neighbors. Thus, the representation $H^{(k+1)}$ at any encoder layer does not use the node's own features $(D_i,X_i)$, but only those of other individuals $(\mathcal{D}_{-i},\mathcal{X}_{-i})$. This is consistent with the interpretation of the exposure $\tilde{Z}_i$, which is intended to summarize how the treatments and characteristics of others affect individual $i$'s outcome.

\section{Testing the validity of exposure mappings}\label{testability}

In this section, we discuss the testability of whether a researcher-defined exposure mapping is correctly specified for capturing all interference effects.
Our approach uses the learned exposures $\tilde{Z}_i$, $i=1,\dots,n$, from Section \ref{exposuremappings} as an instrument to assess whether the (simpler) exposures $\dot{Z}_i$ sufficiently captures all interference. Specifically, if the simpler mapping accurately captures all interference effects, then the exposure $\tilde{Z}_i$ is independent of the individual’s outcome, conditional on $\dot{Z}_i$. By examining violations of this condition, we can assess whether the researcher-defined mapping fails to capture some of the underlying interference.
Notably, the true exposure $Z_i=\mathcal{F}(\mathcal{A},\mathcal{D}_{-i},\mathcal{X}_{-i})$, for $i=1,\dots,n$, is i.i.d. by assumption. We therefore retain the index $i$ in the learned and researcher-defined exposures $\tilde{Z}_i$ and $\dot{Z}_i$ for notational convenience.
Next, we introduce the assumptions that underlie our testing approach for validating the exposure mapping.
These assumptions formalize the structural restrictions on how the exposure variables may depend on one another and on the outcome. We further assume that any statistical independencies correspond to d-separation in the underlying causal model, a condition known as causal faithfulness and formally stated below.
\begin{assumption}[Testing method - causal structure and faithfulness]\label{ass4}
We assume that
$$
\dot{Z}_i(y)=\dot{Z}_i,\textit{ and }\tilde{Z}_i(\dot{z},y)=\tilde{Z}_i \quad \forall \dot{z} \in \mathcal{\dot{Z}} \textit{ and } y \in \mathcal{Y},$$
where $\mathcal{\dot{Z}}$ and $\mathcal{Y}$ denote the corresponding support of $\dot{Z}_i$ and $Y$
and that
\textit{only variables which are d-separated in some causal model are statistically independent}.
\end{assumption}

The first part of Assumption \ref{ass4} rules out reverse causal effects of outcome $Y$ on $\dot{Z}_i$ and $\tilde{Z}_i$ as well as any causal effect of $\dot{Z}_i$ on $\tilde{Z}_i$. The latter reflects the fact that $\tilde{Z}_i$ is a causal parent of $\dot{Z}_i$, which is consistent with the interpretation of $\tilde{Z}_i$ as a more complex mapping from the network structure and neighbor treatments, while $\dot{Z}_i$ represents a simpler transformation thereof.
The following assumption requires that every possible combination of $\tilde{Z}_i$ and $\dot{Z}_i$ occurs with positive probability.
\begin{assumption}[Testing method - common support]\label{ass5}
\begin{align*}
Pr(\dot{Z}_i = \dot{z}, \tilde{Z}_i = \tilde{z})  > 0 \quad \forall \dot{z} \in \mathcal{\dot{Z}}, \tilde{z} \in \mathcal{\tilde{Z}}
\end{align*}
where $\mathcal{\dot{Z}}$ and $\mathcal{\tilde{Z}}$ denote the corresponding support of $\dot{Z}_i$ and $\tilde{Z}_i$.
\end{assumption}

In Assumption \ref{ass6}, we impose that the simpler exposure $\dot{Z}_i$ and the learned exposure $\tilde{Z}_i$ are statistically dependent.
\begin{assumption}[Testing method - dependence between $\dot{Z}_i$ and $\tilde{Z}_i$]\label{ass6}
\begin{align*}
\dot{Z}_i \not\!\perp\!\!\!\perp \tilde{Z}_i. \\
\end{align*}
\end{assumption}

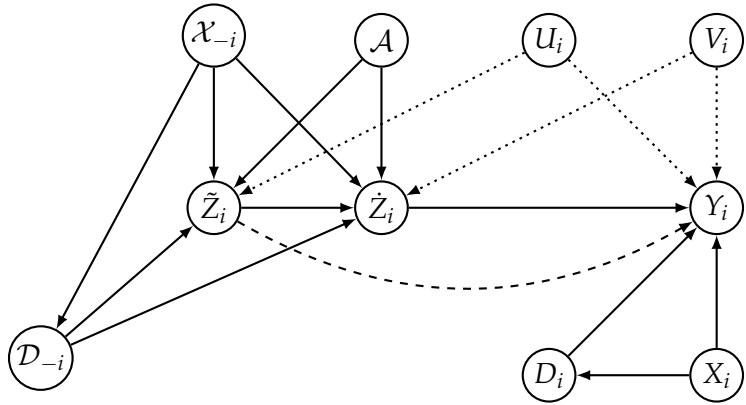
\begin{figure}[ht!]
\centering
\caption{%
Causal diagram underlying the testing method
}\label{fig:dag2}
\begin{tikzpicture}[transform shape, node distance=1.5cm, thick, roundnode/.style={circle, draw, inner sep=1pt, minimum size=7mm}, squarenode/.style={rectangle, draw, inner sep=1pt, minimum size=7mm}]

\node[roundnode] (D) {$\mathcal{D}_{-i}$};
\node[roundnode, right=of D, yshift=2cm] (Z) {$\tilde{Z}_i$};
\node[roundnode, right=of Z] (Zdot) {$\dot{Z}_i$};
\node[roundnode, above=of Zdot] (A) {$\mathcal{A}$};
\node[roundnode, above=of Z] (X) {$\mathcal{X}_{-i}$};
\node[roundnode, right=of A] (U) {$U_i$};
\node[roundnode, right=of U] (V) {$V_i$};
\node[roundnode, below=of V] (Y) {$Y_i$};
\node[roundnode, below=of Y] (Xi) {$X_i$};
\node[roundnode, left=of Xi] (Di) {$D_i$};

\draw[-latex] (D) -- (Z);
\draw[-latex] (D) -- (Zdot);
\draw[-latex] (A) -- (Z);
\draw[-latex] (A) -- (Zdot);
\draw[-latex] (Z) -- (Zdot);
\draw[-latex] (Zdot) -- (Y);
\draw[-latex] (Di) -- (Y);
\draw[-latex] (Xi) -- (Y);
\draw[-latex] (Xi) -- (Di);
\draw[-latex] (X) -- (Z);
\draw[-latex] (X) -- (Zdot);
\draw[-latex] (X) -- (D);
\draw[-latex, dash pattern=on 1pt off 2pt] (U) -- (Z);
\draw[-latex, dash pattern=on 1pt off 2pt] (U) -- (Y);
\draw[-latex, dash pattern=on 1pt off 2pt] (V) -- (Zdot);
\draw[-latex, dash pattern=on 1pt off 2pt] (V) -- (Y);
\draw[-latex, dashed] (Z) edge [bend right=30] (Y);
\end{tikzpicture}

\end{figure}

Together with Assumption \ref{ass4}, which rules out any effect of $\dot{Z}_i$ on $\tilde{Z_i}$, Assumption \ref{ass6} ensures that $\tilde{Z}_i$ causally affects $\dot{Z}_i$.
This corresponds either to a first-stage relationship in the IV literature or to the presence of (potentially unobserved) characteristics that jointly influence both $\dot{Z}_i$ and $\tilde{Z}_i$.
This assumption is satisfied in Figure \ref{fig:dag2}, where $\tilde{Z}_i$ serves as a causal parent of $\dot{Z}_i$.
Assumptions \ref{ass4} and \ref{ass6}, allow us to apply Theorem 1 of \citet{huberkueck2022} in order to construct a test for validating the exposure mapping. It is worth noting that we still rely on the causal structure shown in Figure \ref{fig:dag} and Figure \ref{fig:dag2}. Most importantly, we assume that there are no unobserved confounder jointly affecting $D_i$ and $\tilde{Z_i}$ or $D_i$ and $\dot{Z_i}$.

\begin{theorem}{Conditional on Assumptions \ref{ass4} and \ref{ass6}, it holds that}\label{theorem}
\begin{eqnarray}\label{mainresult}
&Y_i(d,\dot{z}) \perp\!\!\!\perp \dot{Z}_i ,\quad Y_i(d,\dot{z}) \perp\!\!\!\perp  \tilde{Z}_i 
\iff & Y_i \perp\!\!\!\perp \tilde{Z}_i  | \dot{Z}_i=\dot{z}, \quad\forall \dot{z} \in \mathcal{\dot{Z}}, d\in\{0,1\}.
\end{eqnarray}
Conditional on Assumptions \ref{ass4} and \ref{ass6}, the testable implication $Y_i \perp\!\!\!\perp \tilde{Z}_i  | \dot{Z}_i=\dot{z}$ is necessary and sufficient for the joint satisfaction of $Y_i(d,\dot{z}) \perp\!\!\!\perp \dot{Z}_i$ and $Y_i(d,\dot{z}) \perp\!\!\!\perp  \tilde{Z}_i$ when considering potential outcomes $Y_i(d,\dot{z})$ matching $\dot{Z}_i=\dot{z}$ and $D_i=d$.
\end{theorem}

This statement follows directly from Theorem 1 of \citet{huberkueck2022}, considering $\dot{Z}_i$ as the treatment variable and $\tilde{Z}_i$ as the suspected instrument.
The first two conditions of Theorem \ref{theorem} state two independence conditions, which correspond to selection-on-observables for the exposure $\dot{Z}_i$ and instrument validity for the learned exposure $\tilde{Z}_i$, respectively. The first condition imposes that the potential outcome $Y_i(d,\dot{z})$ is independent of $\dot{Z}_i$ for all possible exposure levels $\dot{z} \in \mathcal{\dot{Z}}$ and $d\in\{0,1\}$, i.e., $Y_i(d,\dot{z}) \perp\!\!\!\perp \dot{Z}_i$. This rules out unobserved confounding between $\dot{Z}_i$ and $Y_i$. In Figure \ref{fig:dag2}, this corresponds to the absence of dotted arrows from the unobserved confounder $V_i$ to both $\dot{Z}_i$ and $Y_i$.
The second independence condition of Theorem \ref{theorem} imposes instrument validity for the learned exposure $\tilde{Z}_i$. It requires that $Y_i(d,\dot{z})$ is independent of $\tilde{Z}_i$, i.e., $Y_i(d,\dot{z}) \perp\!\!\!\perp  \tilde{Z}_i$. This means that the learned mapping does not directly affect the potential outcome and that there are no unobserved confounders jointly affecting $\tilde{Z}_i$ and $Y_i(d,\dot{z})$. In Figure \ref{fig:dag2}, this corresponds to the absence of the dotted arrows from the unobserved confounder $U_i$ to both $\tilde{Z}_i$ and $Y_i$.
Therefore, the second condition of Theorem \ref{theorem} ensures that $\tilde{Z}_i$ satisfies the same type of independence requirements as a valid instrument in the IV literature.
Following Theorem 1 of \citet{huberkueck2022}, the testable conditional independence $Y_i \perp\!\!\!\perp \tilde{Z}_i  | \dot{Z}_i=\dot{z}$ in the third condition of Equation \eqref{mainresult} is necessary and sufficient for the joint satisfaction of the first two independence conditions.
In our setting, this provides the following interpretation of the testable implication: we assess whether the learned exposure $\tilde{Z}_i$ contains any residual association with the outcome $Y_i$ after conditioning on the predefined exposure $\dot{Z}_i$. If such an association remains, the mapping $\dot{Z}_i$ fails to capture all relevant interference, and the learned mapping offers additional, causally meaningful information. In turn, if the conditional independence holds, the simpler exposure mapping is sufficient for capturing interference effects. In Figure \ref{fig:dag2} the dashed line (i.e., $\tilde{Z}_i \dashrightarrow Y_i$) indicates a violation of the testable condition, as $\dot{Z}_i$ does not fully capture the interference effects of $\tilde{Z}_i$ on $Y_i$ due to its definition being too simple to account for all forms of interference.

In this paper, we focus on mean conditional independence between the outcome variable and the learned exposure mapping rather than on the full outcome distribution. This is sufficient for identifying average effects, see Theorem 2 of \citet{huberkueck2022}, within our framework of average direct and interference effects defined in Equation \eqref{interferenceeff2}. Modifying the testable implication in Equation \eqref{mainresult} to hold in expectation yields the following testable implication:
\begin{align}\label{meanresult2}
 E[Y_i | \dot{Z}_i = \dot{z}_i, \tilde{Z}_i = \tilde{z}_i]=E[Y_i | \dot{Z}_i = \dot{z}_i] \quad \forall \dot{z} \in \mathcal{\dot{Z}}, \tilde{z} \in \mathcal{\tilde{Z}}.
\end{align}
Equation \eqref{meanresult2} requires that all combinations of $\dot{Z}_i$ and $\tilde{Z}_i$ occurring in the conditioning sets are observed with positive probability. This is ensured by Assumption \ref{ass5}.
Following \citet{huberkueck2022} and \citet{Apfel2023learning}, testing whether the difference in expected outcomes in Equation \eqref{meanresult2} equals zero requires checking this condition for all values of $\dot{Z}$ and $\tilde{Z}$ in their respective supports, which can imply infinitely many testable implications. The solution is to test Equation \eqref{meanresult2} globally, i.e., across all values of $\dot{Z}$ and $\tilde{Z}$, using an aggregated $L_2$-type measure proposed in equation (3.4) in \citet{Apfel2023learning}. Formally, we test $H_0: \theta_0 = 0$ where $\theta_0=(E[Y_i | \dot{Z}_i = \dot{z}_i, \tilde{Z}_i = \tilde{z}_i]-E[Y_i | \dot{Z}_i = \dot{z}_i])^2+(E[Y_i | \dot{Z}_i = \dot{z}_i, \tilde{Z}_i = \tilde{z}_i]-E[Y_i | \dot{Z}_i = \dot{z}_i])$ which evaluates the testable implication in Equation \eqref{meanresult2}.
Relying on the DML approach, \citet{Apfel2023learning} derive an estimator $\hat{\theta}_0$ that is asymptotically normal and $\sqrt{n}$-consistent under suitable regularity conditions. We use this estimator to test whether the researcher-defined exposure is correctly specified, i.e., whether Equation \eqref{meanresult2} holds true. 
The technical details of the testing approach are given in Appendix \ref{appendix_testing}.

\section{Estimation of causal effects}\label{estimation}

In this section, we outline the estimation of the causal effects defined in Section \ref{assumptions}. In the following the exposure mapping $Z_i$ is either the researcher-defined exposure mapping, $\dot{Z}_i$, or the learned exposure mapping, $\tilde{Z}_i$, depending on the result of the testing method introduced in Section \ref{testability}. If $H_0$ is rejected, the learned exposure mapping is used for the estimation of the causal effect, i.e., $Z_i = \tilde{Z_i}$. If $H_0$ is not rejected, the researcher-defined exposure mapping is used instead, i.e., $Z_i = \dot{Z_i}$. When $Z_i = \tilde{Z_i}$, the exposure mapping may be difficult to interpret. Hence, our focus is on the identification and estimation of the direct effect averaged over all exposure levels, $\gamma = E[\gamma(Z)]$. 

To estimate these causal effects, we build on the identification assumptions introduced in Section \ref{assumptions}. While IPW identification is valid under correct specification of the propensity score, outcome-regression-based identification using a model for the conditional mean outcome is valid under correct specification of that model. Both methods can be combined in a doubly robust (DR) approach, which remains consistent if either the propensity score or the outcome model is correctly specified \citep{Robins+94, RoRo95}. Moreover, the DR approach is first-order insensitive to small deviations of both the propensity score and the conditional mean outcome from their respective true models, a property known as Neyman-orthogonality \citep{Neyman1959}. This property is key for the application of the DML framework of \cite{Chetal2018}, in which propensity scores and outcome models are estimated with machine learning methods that may be prone to approximation errors. If the propensity score and outcome models are estimated with convergence rate $o(n^{-1/4})$, DML estimation of average direct, interference and total effects can attain $\sqrt{n}$-consistency under certain regularity conditions. Denoting by $\mu_i(d,z,x) = E[ Y_i | D_i = d, Z_i = z, X_i = x]$ the conditional mean outcome, the DR expression for the mean potential outcome is given by  
\begin{align}\label{drinterference}
E[Y_i(d,z)]&=E \bigg[ \underbrace{\mu_i(d,z,x) +  \frac{I\{D_i=d, Z_i = z\}}{p_i(d,z)} ( Y_i - \mu_i(d,z,x))}_{:=\phi(W_i)}\bigg],
\end{align}
where $\psi(W_i)=\phi(W_i)-E[Y_i(d,z)]$ is the efficient score function.

The outcome regression $\mu_i(d,z,x)$ can be estimated using any machine learning method that satisfies the required convergence rate. The propensity score $p_i(d,z)$ defined in Equation \eqref{pscore} consists of two components. The first component $\Pr(D_i = d | X_i)$ can be estimated using, for example, logistic regression when the treatment is binary. The second component $\Pr(Z_i = z | \mathcal{X}_{-i}, \mathcal{A})$ is conditional on the network $\mathcal{A}$, which is why we suggest estimating it using the GNN-based propensity score estimator introduced in \cite{leung2022graph}. They use a graph isomorphism network (GIN) to estimate the propensity score, that is, the conditional distribution of exposure given the covariates of the other individuals $\mathcal{X}_{-i}$ and the network structure $\mathcal{A}$. Since the exposure can take multiple values, the GIN produces a probability for each possible exposure level through a multiclass output layer. The target variable is a one-hot-encoding of the observed exposure level. 
The model is trained using a logistic loss, $\exp(\hat{m}_z)/\sum_{z' \in \mathcal{Z}}\exp(\hat{m}_{z'})$, to estimate the  propensity score, where $\hat{m}_z$ denotes the GIN output corresponding to exposure level $z$.
They show that, under approximate neighborhood interference (ANI), the propensity score can be estimated with a convergence rate of $o(n^{-1/4})$. This allows valid estimation of the average direct effect, $\gamma = E[\gamma(Z)]$, which we will also show in a simulation study in the next section.

\section{Simulation}\label{sim}

This section provides a simulation study to investigate the finite sample behavior of the testing approach as well as the direct effect estimation. The data are generated according to the following data generating process:
\begin{align*}
Y_i &= \alpha + \delta Z_i + \gamma D_i + \xi X_i + \varepsilon_i, \quad \text{ where } \varepsilon_i \sim N(0,1), \\
D_i &\sim bernoulli\Big(\frac{1}{1+\exp(-(1+2X_i))}\Big), \quad X_i \sim bernoulli(0.5), \\
\quad a_{ij} &= I\{|\rho_i - \rho_j | \leq r_n\} \text{ with } \rho_i \sim U([0,1])^2 \text{ and } r_n = \Big(\frac{30}{\pi n} \Big)^{(1/2)},
\end{align*}
where $(\alpha, \delta, \gamma, \xi) = (-1, 5, 1, 1)$. The outcome $Y_i$ is a linear function of the individual treatment $D_i$, the true exposure $Z_i$, the covariate $X_i$ and an unobservable $\varepsilon_i$. The covariate $X_i$ is a binary variable, and the binary treatment $D_i$ is a function of $X_i$. Following \citet{leung2022graph}, the network structure $\mathcal{A}$ is generated from a random geometric graph model. The GCA used for learning the exposure mapping consists of two graph convolutional layers in the encoder and one regression-based decoder. The learning rate is 0.01 and the number of epochs is 200. For estimating the score function, the support of the continuous learned exposure variable is partitioned based on the quartiles of its distribution, i.e., $L=4$.

We consider three settings to assess the performance of our testing approach, using $200$ Monte Carlo replications for sample sizes $n = 500, 1000, \text{ and } 2000$. In the first setting, the true exposure mapping is given by the share of treated neighbors weighted by their covariates, i.e., $Z^{S1}_i =\frac{\sum_{j \ne i}a_{ij}D_j X_j}{\sum_{j \ne i}a_{ij}}$, and the researcher-defined exposure mapping coincides with the true exposure, i.e., $\dot{Z}^{S1}_i =\frac{\sum_{j \ne i}a_{ij}D_j X_j}{\sum_{j \ne i}a_{ij}}$. Thus, the researcher-defined exposure mapping is correctly specified and the exposure learned by the GCA should not contain additional information beyond the researcher-defined mapping. In Setting 2, the true exposure mapping depends on both first-order and second-order network neighborhoods. Specifically, the true exposure is given by $Z^{S2}_i = \frac{\sum_{j \ne i}a_{ij}D_j X_j}{\sum_{j \ne i}a_{ij}} + \frac{\sum_{k \ne i}b_{ik}D_k X_k}{\sum_{k \ne i}b_{ik}}$, where $b_{ik} := I\{\sum_{i \ne j, j \ne k} a_{ij}a_{jk} > 0\} \cdot (1-a_{ik}) \cdot I\{k \ne i \}$. The first term captures again the share of treated neighbors weighted by their covariates. The second term of the true exposure captures second-degree neighbors of individual $i$, i.e., nodes that are connected to $i$ via a path of length two, excluding $i$ itself and all direct neighbors. In contrast, the researcher-defined exposure is binary and only accounts for direct neighbors, i.e., $\dot{Z}^{S2}_i =I\{ \sum_{j \ne i} a_{ij}D_j X_j > 0\}$. Thus, the researcher-defined exposure is misspecified. In the third setting, the researcher-defined exposure is equal to the one in Setting 1, i.e., $\dot{Z}^{S1}_i=\dot{Z}^{S3}_i$. However, the true exposure is a nonlinear transformation of the cumulative treated neighborhood intensity with an explicit threshold and saturation: $Z^{S3}_i=1- \exp \Big(-0.5 \; \max\!\left\{0, \;\sum_{j \ne i}a_{ij}D_j X_j - 10 \right\} \Big)$. This setting therefore also represents a case of misspecification if researchers were to apply a linear specification to model the conditional mean outcome in Equation \eqref{drinterference} based on $\dot{Z}^{S3}_i$.

\begin{table}[h]
\centering
\small
\begin{tabular}{c | cc | cc | cc}
\toprule\toprule
& \multicolumn{2}{c}{Setting 1} & \multicolumn{2}{c}{Setting 2} & \multicolumn{2}{c}{Setting 3} \\
\cmidrule(lr){2-3}\cmidrule(lr){4-5}\cmidrule(lr){6-7}
$n$ & rejection rate & mean p-value & rejection rate & mean p-value & rejection rate &mean p-value \\
\midrule
500  & 0.00 & 0.76 & 0.32  & 0.25 &0.8  &0.11 \\
1000 & 0.00 & 0.73 & 0.625 & 0.10 &  0.805 &  0.09\\
2000 & 0.00 & 0.69 & 0.895 & 0.03 & 0.83 &  0.09 \\
\bottomrule\bottomrule
\end{tabular}
\parbox{\textwidth}{\centering \footnotesize
Notes: 'rejection rate' gives the empirical rejection rate when setting the level of statistical significance to 0.05 (or 5\%); R = 200  replications.}
\caption{Simulation results - testing method}\label{sim:test}
\end{table}

Table \ref{sim:test} reports the empirical rejection rates and mean p-values of the testing approach across the three settings. In Setting 1, where the researcher-defined exposure mapping is correctly specified, the test never rejects the null hypothesis and the p-values remain high across all sample sizes. In Settings 2 and 3, where the researcher-defined exposure mapping is misspecified, the rejection rate increase with the sample size, while the mean p-values decrease and approach the significance level $\alpha = 0.05$. Overall, these results are consistent with the theoretical implications discussed in Section \ref{testability}. 

In the second part of the simulation study, we asses the estimation of the direct effect based on the learned exposure $\tilde{Z}_i$ obtained from a GCA. The direct effect is estimated using IPW, where the propensity score for the treatment assignment, $Pr(D_i =1 | X_i)$, is estimated via logistic regression, and the exposure propensity $Pr(Z_i | \mathcal{X}_{-i}, \mathcal{A})$ is approximated using an oracle estimator based on the data-generating process. The true exposure that the GCA aims to recover is defined as $Z_i =I\{ \sum_{j \ne i} a_{ij}D_j X_j > 2\}$ and we adjust the parameter for generating $a_{ij}$ to $r_n = \Big(\frac{5}{\pi n} \Big)^{(1/2)}$ in the DGP outlined in the main text.

\begin{table}[h]
\centering
\small
\begin{tabular}{ c | c  c  c } 
\toprule\toprule
$n$ & est & std & bias \\
\midrule
100 & 1.178 & 1.047 & 0.843 \\
200 & 1.128 & 0.684 & 0.542 \\
500 &  1.007 & 0.396 & 0.314 \\
1000 & 1.005&  0.306  &  0.241 \\
\bottomrule\bottomrule
\end{tabular}
\parbox{\textwidth}{\centering \footnotesize Notes: Average estimate of $\gamma$ (est), its standard deviation (std), and the average absolute estimation error (bias) across $R=200$ replications. The true value of the direct effect is $\gamma = 1$.}
\caption{Simulation results - direct effect estimation}\label{sim:dir}
\end{table}

Table \ref{sim:dir} reports the simulation results for the estimation of the direct effect $\gamma$ based on the learned exposure $\tilde{Z}_i$. For small sample sizes, the estimator shows noticeable variability and bias. As the sample size increases, both the bias and the standard deviation decrease. For sample sizes $n=500$ and $n=1000$, the average estimated direct effect is close to the true value $\gamma = 1$. This indicates improved precision and convergence toward the true direct effect.

\section{Conclusion}\label{conclusion}

In this paper, we develop a data-driven approach to learn exposure mappings in the presence of interference, instead of relying on a priori defined mapping. We use a graph convolutional autoencoder to learn exposure mappings that summarize how others' treatment assignments affect an individual's outcome. Since the identification of average direct effects depends crucially on the correct specification of the exposure mapping, we study whether a simple, researcher-defined exposure mapping is sufficient or a more complex, learned mapping is required. To this end, we propose a testing method based on conditional independence implications. This test evaluates whether a researcher-defined exposure mapping captures all relevant interference. Violations of the testable implication indicate that the predefined exposure mapping is misspecified and that a learned mapping should be used for estimating the direct effect. Overall, our study provides guidance on how to learn and validate exposure mappings in the presence of interference.

\clearpage

\newpage

 \bibliography{lit.bib}

{\large \renewcommand{\theequation}{A-\arabic{equation}}
\setcounter{equation}{0} \appendix }
\appendix \numberwithin{equation}{section}

\begin{appendix}

\section{Graph convolutional autoencoder}\label{GCA}

\begin{figure}[!h]
\centering
\includegraphics[scale=1.0]{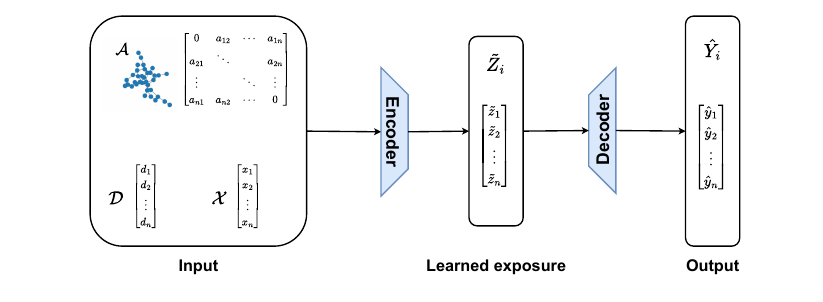}
\caption{Graph convolutional autoencoder architecture}
\label{fig:architecture}
\end{figure}

\section{Details on the testing approach described in Section \ref{testability}}\label{appendix_testing}
In this section, we outline the technical details of our testing approach, which builds on the work of \citet{huberkueck2022} and \citet{Apfel2023learning}.
We test whether the difference in expected outcomes in Equation \eqref{meanresult2} equals zero using an aggregated $L_2$-type measure. Formally, we want to test $H_0: \theta_0 = 0$ where 
$$\theta_0=(E[Y_i | \dot{Z}_i = \dot{z}_i, \tilde{Z}_i = \tilde{z}_i]-E[Y_i | \dot{Z}_i = \dot{z}_i])^2+(E[Y_i | \dot{Z}_i = \dot{z}_i, \tilde{Z}_i = \tilde{z}_i]-E[Y_i | \dot{Z}_i = \dot{z}_i]).$$ 
Given that the learned $\tilde{Z}$ is continuous in the proposed architecture in Section \ref{exposuremappings}, we need to discretize its support. Let $l=1,\dots,L$ denote a partition of its support $\mathcal{\tilde{Z}}$ with $\cup_{l}\tilde{Z}_l=\mathcal{\tilde{Z}}$.
Applying the work of \cite{huberkueck2022} to our context, let $\mu( \dot{Z}_i, \tilde{Z}_i\in\tilde{Z}_{l}):=E[Y_i| \dot{Z}_i, \tilde{Z}_i\in\tilde{Z}_{l}]$, $p_l(\dot{Z}_i)=\Pr(\tilde{Z}_i \in \tilde{Z}_l | \dot{Z}_i)$ and $1(\tilde{Z}_i\in \tilde{Z}_l)$ be an indicator function, which takes the value one if $\tilde{Z}_i$ falls into the partition $\tilde{Z}_l$ (and zero otherwise).
Therefore, we test the following null hypothesis $H_0$:
\begin{align*}
    \widetilde\theta:=  E\left[\sum_{l=1}^L[(\mu(\dot{Z}_i,\tilde{Z}_i\in \tilde{Z}_l)-\mu(\dot{Z}_i,\tilde{Z}_i\notin \tilde{Z}_l))^2+(\mu(\dot{Z}_i,\tilde{Z}_i\in \tilde{Z}_l)-\mu(\dot{Z}_i,\tilde{Z}_i\notin \tilde{Z}_l))]\right] = 0.
\end{align*}
\citet{Apfel2023learning} propose the following Neyman-orthogonal score for testing
\begin{align}\label{score4}
&\quad \psi(W_i,\theta,\eta)\\
&=\sum_{l=1}^L(\mu(\dot{Z}_i,\tilde{Z}_i\in \tilde{Z}_l)-\mu(\dot{Z}_i,\tilde{Z}_i\notin \tilde{Z}_l))^2\notag\\&
+\sum_{l=1}^L2(\mu(\dot{Z}_i,\tilde{Z}_i\in \tilde{Z}_l)-\mu(\dot{Z}_i,\tilde{Z}_i\notin \tilde{Z}_l))\notag\\
&\left(\frac{(Y_i-\mu(\dot{Z}_i,\tilde{Z}_i\in \tilde{Z}_l)) 1(\tilde{Z}_i\in \tilde{Z}_l)}{p_l(\dot{Z}_i)}
-\frac{(Y_i-\mu(\dot{Z}_i,\tilde{Z}_i\notin \tilde{Z}_l))1(\tilde{Z}_i\notin \tilde{Z}_l)}{1-p_l(\dot{Z}_i)}\right)\notag\\&
+\sum_{l=1}^L(\mu(\dot{Z}_i,\tilde{Z}_i\in \tilde{Z}_l)-\mu(\dot{Z}_i,\tilde{Z}_i\notin \tilde{Z}_l))\notag\\&
+\sum_{l=1}^L\left(\frac{(Y_i-\mu(\dot{Z}_i,\tilde{Z}_i\in \tilde{Z}_l)) 1(\tilde{Z}_i\in \tilde{Z}_l)}{p_l(\dot{Z}_i)}
-\frac{(Y_i-\mu(\dot{Z}_i,\tilde{Z}_i\notin \tilde{Z}_l)) 1(\tilde{Z}_i\notin \tilde{Z}_l)}{1-p_l(\dot{Z}_i)}\right)\notag - \theta
\end{align}

and show that using this score function within the DML framework  \citep{Chetal2018}, leads to an estimator $\hat{\theta}_0$ which is asymptotically normal and $\sqrt{n}$-consistent under suitable regularity conditions. Especially, we need to ensure that the nuisance functions $\mu(\cdot)$ and $p_l(\cdot)$ can be estimated at rate $o(n^{-1/4})$.
Hence, we can use this DML estimator to test whether the researcher-defined exposure is correctly specified, i.e., whether Equation \eqref{meanresult2} holds true.

\end{appendix}

\end{document}